 \documentclass[final,twocolumn]{elsarticle}
\usepackage{algpseudocode,algorithm}
\algrenewcommand\textproc{\texttt} 

\usepackage{multirow}
\usepackage{tikz}
\usepackage{verbatim}
\usetikzlibrary{arrows}
\usetikzlibrary{calc}

\usepackage{physics}
\usepackage{amsmath}
\usepackage{graphicx}
\usepackage{braket}
\usepackage{comment}
\usepackage{hyperref}
\usepackage{soul,color}
\usepackage{bbold}
\usepackage{xcolor}

\newcommand{\symset}{\mathcal{S}}
\newcommand{\cntset}{\mathcal{P}} 
\newcommand{\indset}{\mathcal{I}}

\newcommand{\maptop}{{{\mathcal{M}}}_{p-1 \mapsto p}}

\newcommand{\com}[1]{}

\newcommand{\maptoN}{{{\mathcal{M}}}_{N-1 \mapsto N}}

\graphicspath{{./pdf/}}
\usepackage{float}

\usepackage{amssymb}
\newcounter{bla}

\journal{Computer Physics Communications}

\begin{document}

\begin{frontmatter}

\title{Efficient linear scaling mapping for permutation symmetric Fock spaces}
\author{M. Ahsan Zeb \corref{author}}
\date{\today}
\cortext[author]{Corresponding author.\\\textit{E-mail address:} ahsan.zeb@hotmail.com}
\address{Department of Physics, Quaid-i-Azam University, Islamabad 45320, Pakistan}

\begin{abstract}
Numerically solving a second quantised many-body model in the permutation symmetric Fock space
can be challenging for two reasons: 
(\emph{i}) an increased complication in the calculations of the matrix elements of various operators,
and (\emph{ii}) a poor scaling of the cost of these calculations with the Fock space size.
We present a method that solves both these problems.
We find a mapping that can be used to 
simplify 
the calculations of the matrix elements.
The mapping is directly generated so its computational cost scales only linearly with the space size and is negligible even for large enough sizes that approach the thermodynamic limit.
A fortran implementation of the method as a library -- FockMap --
is provided along with a test program.

\com{ Major Changes:
--- program changed into a library.
--- intro: extended introduction. two paragraphs,
--- program section, completely rewritten, includes a few exemplary models.
--- conclusions: rewritten, includes take home message, limitations, possible future developments.
}
\end{abstract}

\twocolumn


\end{frontmatter}

{\bf PROGRAM SUMMARY}

\begin{small}
\noindent
{\em Program Title: FockMap}                                          \\
{\em Licensing provisions: GPLv3}                                   \\
{\em Programming language: FORTRAN}                                   \\
{\em Nature of problem: Solving second quantised many-body models in permutation symmetric Fock space}\\
{\em Solution method:
A mapping between the Fock states exists that can be used to calculate the matrix elements of various operators. The pattern in the mapping is found and directly generated.
 }\\
 \end{small}

In quantum optics, condensed matter and related fields, e.g., atomic and molecular physics, quantum information and quantum computing, 
the systems we study are often composed of or contain
one or more types of 
many \emph{identical} subsystems.
These many-body problems cannot be reduced to simpler single-body
problems due to the presence of the interactions between the
subsystems. 
Some common examples of such subsystems include
electronic systems like
multi-level atoms, molecules or quantum dots, 
phononic systems like
quantised intramolecular or lattice vibrational modes,
and
photonic systems like
quantised microcavity modes.
When the interactions between the subsystems 
are strong, 
the standard perturbative approaches cannot be used so
we resort to non-perturbative methods like exact diagonalisation.

Such models are often cast in the second quantised form
and solving them numerically in their Fock space 
requires computation of matrix elements of the Hamiltonian and other operators. 
In this regard,
a particular indexing system for the basis states can be more efficient than others~\cite{streltsov2010}.
Since the size of the Hilbert space of a many-body system
scales exponentially with the system size,
the computational cost becomes too large 
at relatively very small system sizes, far away from the desired thermodynamic limit.
This bad scaling can sometimes be improved by working only in a suitable subspace of the Hilbert/Fock space.
A particular example of such a case is when 
the Hamiltonian is invariant under the permutation
of the identical subsystems.
Under this condition,
the ground state as well as some other properties 
can be determined 
by working only within the permutation symmetric subspace.
Here, the permutation is considered over the subsystems.
For example, if we consider the Holstein-Tavis-Cummings model to calculate the condensate state of organic polaritons or the absorption spectrum of an organic microcavity,
the subspace that is symmetric under the permutations
of the (identical) excitons and the (identical) phonon modes
suffices~\cite{zeb2020}.

The main advantage of using the permutation symmetry in such a case is that the size of the permutation symmetric subspace scales only polynomially with the system size.
However, there is a cost to pay:
the computation of matrix elements becomes 
non-trivial.
Besides, a brute force calculation of these still scales as
the square of the subspace size, which in return limits the size of the systems that can be dealt with this method. 
We find that this problem boils down to the calculation of a certain mapping.
We present an efficient method to calculate this mapping.
Our method is based on the recognition of the pattern that this mapping acquires when the indexing of the basis states follows a certain order.
This method can be adopted to solve \emph{any} second quantised model in the permutation symmetric Fock space whenever the large system sizes are to be studied.
For example, in Ref.~\cite{zeb2020}, we apply this method on a complex many-body problem with three types of excitations.

Here we illustrate the method for  
a generic many-body system that includes $N$ identical bosonic modes.
In Sec.~\ref{sec:basis-set}, we introduce the permutation symmetric subspace of the Fock space and a specific indexing scheme for its states.
Section~\ref{sec:map} defines the mapping discussed above
and 
Sec.~\ref{sec:matelem} gives two examples of how this mapping can be used
to calculate the matrix elements of operators.
We describe the pattern this mapping acquires 
and a possible algorithm to generate it in Sec.~\ref{sec:compmap}.
At the end, in Sec.~\ref{sec:code}, we describe ``FockMap'' library, a fortran implementation of our method 
that can be used to 
efficiently calculate the mapping and some other properties of the basis states for a given number of identical modes.

\section{Permutation symmetric Fock space}
\label{sec:basis-set}

Consider the Fock space of $N$ identical boson modes,
e.g., $N$ identical harmonic oscillators.
We can make subsets of the Fock states such that each subset contains the states that are related by permutation of their occupation numbers.
If we denote $\{\nu\}$ as the set of the occupation numbers for such a subset, then the
permutation symmetric superposition of the states in it
$\symset_{N}(\{\nu\})$,
is given by,
\begin{equation}
\ket{\symset_{N}\{\nu\}} \equiv 
\frac{\sum_{\text{P}} \text{P}[\ket{\{\nu\}}]}
{\sqrt{\cntset_N(\{\nu\})}},
\end{equation}
where the sum over P indicates a sum over the permutations,
and
$\cntset_N(\{\nu\})$ counts the number of distinct permutations or the size of the subset, which will depend on the pattern of occupations in $\{\nu\}$. 
\emph{The set of all possible permutation symmetric states $\{\symset_{N}(\{\nu\})\}$ spans the permutation symmetric subspace of the Fock space.}
If we label the frequency $f_{\nu_i}$ as the number of times each value $\nu_i$ appears in the set $\{\nu\}$, then the number of permutations is the multinomial coefficient  
\begin{eqnarray}
\label{cntset}
\cntset_N(\{\nu\}) = N!/(\prod_i f_{\nu_i}!).
\end{eqnarray}
For example, 
for the set of occupations $\{0112\}$, the frequencies are $1,2,1$ and so
$\cntset_4(\{0112\})=12$,
and the permutation symmetric state is:
\begin{displaymath}
\ket{\symset_{4}\{0112\}} \equiv 
\frac{
\begin{aligned}
\bigl(&\ket{0112} + \ket{1012} + \ket{1102} + \ket{1120} \\ 
+& \ket{0211}  + \ket{2011}  + \ket{2101}  + \ket{2110} \\ 
+& \ket{0121}+ \ket{1021}+ \ket{1201}+ \ket{1210} \bigr)
\end{aligned}
}{\sqrt{12}}.
\end{displaymath}
To perform numerical calculations, the occupation numbers need to be
truncated.  That is, we need to introduce a cutoff $M$, such that
$\nu_{i}\in [0,M]$. 
In any model, $M$ needs to be sufficiently large for the results to be converged and hence reliable.
The total number of distinct permutation symmetric states
for $N$ modes
is ${}^{M+N}C_{M}$ compared to a total of $(M+1)^N$ states,
which increases only polynomially with $N$, much slower than the exponential scaling of the full Hilbert space.
The counting comes from the number of ways to pick $N$ numbers in the range $[0,M]$ ignoring order.
This far better scaling makes it possible 
to calculate the lowest energy eigenstate and some other properties for large values of $N, M$ to see the behaviour of the model under study in the thermodynamic limit.

\subsection{Indexing the permutation symmetric Fock states}
\label{sec:indexing}
To use the 
the permutation symmetric \emph{basis} states $\{\symset_{N}(\{\nu\})\}$
on computer, 
a suitable indexing is required. That is, an integer $\indset_N(\{\nu\})$ for every $\symset_{N}(\{\nu\})$.
There is no unique way to do it, however, there are important advantages
if we choose to \emph{lexicographically} order the occupations in the set $\{\nu\}$
and index the basis states in order of increasing the occupation from left to right,
as shown in Table~\ref{sets}.

\begin{table}[htp]
\caption{Indexing of the basis states with lexicographically ordered sets of occupation numbers, illustrated for $N=5, M=2$.}
\begin{center}
\begin{tabular}{|c|c|}
\hline
 $\{\nu\}$ & $\indset_N(\{\nu\})$\\
\hline
\{0,0,0,0,0\} & 0 \\
\hline
\{0,0,0,0,1\} & 1\\
\hline
\{0,0,0,0,2\} & 2\\
\hline
\{0,0,0,1,1\}& 3 \\
\hline
\{0,0,0,1,2\} & 4\\
\hline
 \{0,0,1,1,1\}& 5 \\
\hline
 \{0,0,1,1,2\} &6\\
\hline
\{0,0,1,2,2\} & 7\\
\hline
\ldots & \ldots \\
\hline
\{2,2,2,2,2\}& 21 \\
\hline
\end{tabular}
\end{center}
\label{sets}
\end{table}%

\section{Mapping for the permutation symmetric Fock states}
\label{sec:map}
The downside of using the permutation symmetric space is that
the calculation of the matrix elements of the Hamiltonian 
and other operators becomes 
non-trivial. 
To this end, as we show in the following section (\ref{sec:matelem}), 
we find that 
we can always depend on 
a (many-to-one) mapping $\maptoN$ from the permutation symmetric states of $N-1$ modes to those of $N$ modes.
\emph{The mapping $\maptoN$ takes us from the index $\indset_{N-1}(\{\mu\})$ and the occupation of a single mode $n$ (which can itself be treated as an index) to $\indset_N(\{\nu\})$ if adding $n$ to the set $\{\mu\}$ makes the set $\{\nu\}$.}
For the indexing described in sec.~\ref{sec:indexing},
not only $\indset_{N-1}(\{\mu\})$ but also
$\maptoN$ turns out to work for all possible cases we could be interested in with permutation symmetric states of $N,N-1,N-2,...,1$ modes.
That is, the two dimensional integer array on computer for $\maptoN$ can also be used for 
$\maptop$ with $p=1,2,...,N$.

For $N$ modes,
a naive calculation of $\maptoN$ should scale as the square of the size $\mathcal{N}_N$ of the permutation symmetric Fock space, i.e., as $\mathcal{O(N}_N^2)$,
as there are $\mathcal{N}_N \times M$ integers 
(in the space of $N$ modes)
to compare to
$\mathcal{N}_{N-1} \times M \times M$ integers.
(Some restrictions can be imposed to improve the scaling though.)
But, thanks to our indexing, $\maptoN$ inherits a pattern that can be easily recognised and generated directly thus completely avoiding this bottleneck.
This direct generation of $\maptoN$ is not only an $\mathcal{O(N})$ process, it has a very small prefactor such that
the computational cost even for $\mathcal{N}\sim 10^9$ is 
negligible.

\section{Using the Mapping: Matrix elements of operators}
\label{sec:matelem}

Let's calculate the matrix elements of a few operators
to illustrate how the mapping $\maptoN$ can be used.

\subsection{Reduced density matrix of a single mode} 
\label{sec:rho}
In this section, we discuss how the mapping $\maptoN$
can help us determine the reduced density matrix. 
We can write eigenstate as follows:
\begin{equation}
  |\Psi\rangle = 
  \sum_{\substack{k_N\equiv\indset_N{(\{\nu\})}}}\!\!\!\!
  \psi_{k_N}  \ket{\symset_{N}\{\nu\}},
\end{equation}
where $\{\psi_{k_N} \}$ is the array of coefficients that is computed.
A crucial step to calculating observables is to define the reduced density matrix $\rho^{r}$
that describes the (mixed) state of a single mode. It requires taking a trace over the states of all modes other than the one in question.  This can be written as:
\begin{equation}
  \rho^{r}_{m,m^\prime} \equiv
  \braket{\Psi| m^\prime}
  \braket{m |\Psi}
\end{equation}
Here $i$ is an arbitrary mode (since states are permutation symmetric), and $m, m^\prime$ denote occupation number states on the molecule in question.

To find the element ${\rho^r}_{m,m^\prime}$, 
we need to trace out the states of the $N-1$ other modes
and hence 
find all pairs
of states with $N$ excited modes 
which are reduced to the same $N-1$
mode state when $m,m^\prime$ are taken out. 
If we denote
$k^{}_{N} = \indset_{N}(\{\nu\})$
and 
$k^{\prime}_{N} = \indset_{N}(\{\nu^\prime\})$
 as the indices of a pair of states $\{\nu\}$ and $\{\nu^\prime\}$ of $N$ modes
 that reduce to the same state $\{\nu^{\prime\prime}\}$ of $N-1$ modes with index
$j^{}_{N-1} = \indset_{N-1}(\{\nu^{\prime\prime}\})$,
we can write
\begin{equation}
  \label{eq:define_p_map}
  \begin{gathered}
  k^{}_N    = \maptoN(m,      j_{N-1}), \\
  k^\prime_N = \maptoN(m^\prime,j_{N-1}).
  \end{gathered}
\end{equation}
With these maps,  we can then trace over $j_{N-1}$, describing the state of
the other modes.
  Assuming the above relations between $j_{N-1}, k^{}_N$,   and $k^\prime_N$, the reduced density matrix takes the form:
\begin{align}
  \label{eq:dm}
  {\rho^r}_{m,m^\prime} &=
  \sum_{j_{N-1}=1}^{\mathcal{N}_{N-1}}
  \frac{
  \psi_{k_N} \psi^{\ast}_{k^\prime_N}
   \cntset_{N-1}(j_{N-1})
  }{\sqrt{\cntset_{N}(k_{N}) \cntset_{N}(k^\prime_{N})}}.
\end{align}
Here $\mathcal{N}_{N-1}$ is the total number of the permutational
symmetric Fock states
involving $N-1$ modes. 
The factors in the denominator come from the normalization of the
permutation symmetric basis states, 
whereas the factor $\cntset_{N-1}(j_{N-1})$ counts the number of matching terms
in the permutation symmetric superposition of the Fock states
$k_{N},k^\prime_{N}$ --- so give unit overlap --- after taking out the
states of our subject molecule. 
Using Eq.~\ref{cntset} and keeping in mind the relationship between
states  $j_{N-1}, k^{}_N,k^\prime_N$, we can simplify Eq.~\ref{eq:dm} to
\begin{align}
  \label{eq:dm1}
  {\rho^r}_{m,m^\prime}
  &=\sum_{j_{N-1}=1}^{\mathcal{N}_{N-1}}
  \frac{\psi_{k_N} \psi^{\ast}_{k^\prime_N}}
  {N \sqrt{f_{m}(\{\nu\}) f_{m^\prime}(\{\nu^\prime\})}}.
\end{align}
Algorithm~\ref{alg:rho} (with ${f_{m,k_N}\equiv f_{m}(\{\nu\})}$, etc.) summarises this computation.

\begin{algorithm}
  \caption{Matrix elements of $\rho^r$}\label{alg:rho}
  \begin{algorithmic}
  \Function{getrho}{}
\For{\texttt{$j_{N-1}$ in range(0,$\mathcal{N}_{N-1}-1$)}} 
     \For{\texttt{$m$ in range(0,M)}} 
     \For{\texttt{$m^\prime$ in range(0,M)}} 
        \State \texttt{$k_N = \maptoN(m,j_{N-1})$} 
        \State \texttt{$k^\prime_N= \maptoN(m^\prime,j_{N-1});$}
	 \State \texttt{$x=\frac{\psi_{k_N} \psi^{\ast}_{k^\prime_N}}
  {N \sqrt{f_{m,k_N} f_{m^\prime,k^\prime_N}}}$ }
         \State \texttt{$\rho^r_{m,m^\prime}=\rho^r_{m,m^\prime}+x$} 
        \EndFor  
         \EndFor  
        \EndFor 
       \State \Return $\rho^r$ 
    \EndFunction
  \end{algorithmic}
\end{algorithm}

\subsection{Creating a delocalised excitation, $\sum_i^N \hat b^\dagger_i$}
\label{sec:matelemb}

Consider the operator
$\sum_i^N \hat b^\dagger_i$.
The matrix element can be written explicitly as a sum over permutations:
\begin{multline} 
\bra{\{\nu^\prime\}_N}
  \sum_{j=1}^N\hat b^\dagger_j
 \ket{\{\nu \}_N} 
 \\=
 \sum_{P^\prime} 
 \frac{P[\bra{\nu^\prime_1\nu^\prime_2\ldots \nu^\prime_N}]}{\cntset_N(\{\nu^\prime\})}
 \sum_{j=1}^N \hat b^\dagger_j
 \sum_{P}\frac{P[\ket{\nu_1\nu_2\ldots \nu_N}]}{\sqrt{\cntset_N(\{\nu\})}}.
 \end{multline}
Consider
$ \sum_{j=1}^N \hat b^\dagger_j
 \sum_{P}P[\ket{\nu_1\nu_2\ldots \nu_N}]$.
Each permutation 
gives terms such as
$\sqrt{\nu_1+1} $ times the state with $\nu_1\to \nu_1 +1$. In general this leads to overlaps of the form:
\begin{align*} 
\sqrt{\nu_i+1}\times P[\bra{\nu^\prime_1\nu^\prime_2\ldots \nu^\prime_N}]  P[\ket{\nu_i\to \nu_i+1 \text{~in~} \{\nu\}_N}],
\end{align*}
which are non-zero only if
$\{\nu^\prime\}$ 
is the same as $\{\nu\}$ 
except $\nu_i\to\nu_i + 1$.
\com{ ---i.e., the \emph{multiset differences} are
$\{\nu \}_N \setminus \{\nu^\prime \}_N = \{\nu_i\}$
and 
$\{\nu^\prime \}_N \setminus \{\nu \}_N=\{\nu_i+1\}$.
}
In other words, the only difference between these two states is
that their frequencies of $\nu_i$ and $\nu_i+1$ are different but still related by
\begin{subequations}
\label{fcond}
\begin{eqnarray}
f_{\nu_i}(\{\nu\})&=&f_{\nu_i}(\{\nu^\prime\}) + 1,\\
f_{\nu_i+1}(\{\nu\})&=&f_{\nu_i+1}(\{\nu^\prime\}) - 1.
\end{eqnarray}
\end{subequations}

If so, every ket in the permutations
finds its dual.
Since,
there are
$\cntset_p(\{\nu\}) $ permutations of
$\{\nu\}$,
 we will get
$ \cntset_p(\{\nu\})\times \sqrt{\nu_i+1}$ for one such term.
The element $\nu_i$ may occur multiple times in the set $\{\nu\}$;
we denote the frequency with which it occurs as $f_{\nu_i}(\{\nu\})$.
The matrix element then becomes
\begin{multline} 
\bra{\{\nu^\prime\}_N}
  \sum_{j=1}^N\hat b^\dagger_j
  \ket{\{\nu \}_N}
\\=
\sqrt{
\frac{\cntset_N(\{\nu\})}
{\cntset_N(\{\nu^\prime\})}
}
f_{\nu_i}(\{\nu\})
\sqrt{\nu_i+1}
\\=
\sqrt{(\nu_i+1)f_{\nu_i}(\{\nu\})(f_{\nu_i+1}(\{\nu\})+1)},
\end{multline}
where we have used the definition of ${\cntset_N(\{\nu\})}$, Eq.~\ref{cntset}.

To use the above expression in a computer program,
we need to know 
the indexes 
$\indset_N(\{\nu\})$ and $\indset_N(\{\nu^\prime\})$
of the two
states $\ket{\symset_{N}\{\nu\}}$ and $\ket{\symset_{N}\{\nu^\prime\}}$
involved.
To determine that we can use the conditions in Eq.~\ref{fcond}.
Suppose we have already indexed states
as in Sec.~\ref{sec:indexing}
and calculated the frequencies $f_{\nu_i}(\{\nu\})$ (and ${\cntset_N(\{\nu\})}$, etc.).
Now, here comes the fun part. 
Eq.~\ref{fcond} can be satisfied if we
start with a given permutation symmetric state 
$\ket{\symset_{N-1}\{\nu''\}}$
of $N-1$ modes with $\{\nu''\}$ being the common subset of the two
sets $\{\nu\},\{\nu^\prime\}$, which can in fact be \emph{any} set!
The mapping $\maptoN$ takes us from the index $\indset_{N-1}(\{\nu''\})$ and $\nu_i$ to $\indset_N(\{\nu\})$.
Similarly, it gives us $\indset_N(\{\nu^\prime\})$
from $\indset_{N-1}(\{\nu''\})$ and $\nu_i+1$.
In summary, Algorithm~\ref{alg:h} is quite
an efficient way to calculate these matrix elements.

\begin{algorithm}
  \caption{Matrix elements of $\hat H = \sum_i\hat b_i$}\label{alg:h}
  \begin{algorithmic}
  \Function{getH}{}
\For{\texttt{$j_{N-1}$ in range(0,$\mathcal{N}_{N-1}-1$)}}  
     \For{\texttt{$m$ in range(0,M)}} 

        \State \texttt{$k_N = \maptoN(m,j_{N-1})$} 
        \State \texttt{$k^\prime_N= \maptoN(m+1,j_{N-1});$} 
	 \State \texttt{$x=\sqrt{(m+1)f_{m,k_N}(f_{m+1,k^\prime_N}+1)}$} 
         \State \texttt{$H_{k_N,k^\prime_N}=H_{k_N,k^\prime_N}+x$} 
        \EndFor  
        \EndFor 
       \State \Return $H$ 
    \EndFunction
  \end{algorithmic}
\end{algorithm}

\section{Calculating the mapping $\maptoN$}
\label{sec:compmap}
In the follwing, we first describe the pattern that the mapping $\maptoN$ acquires
if we use the indexing scheme presented in sec.~\ref{sec:indexing}.
A simple method to generate $\maptoN$ will be presented afterwards.

\begin{figure}[h]
\centering
\includegraphics[width=0.6\columnwidth]{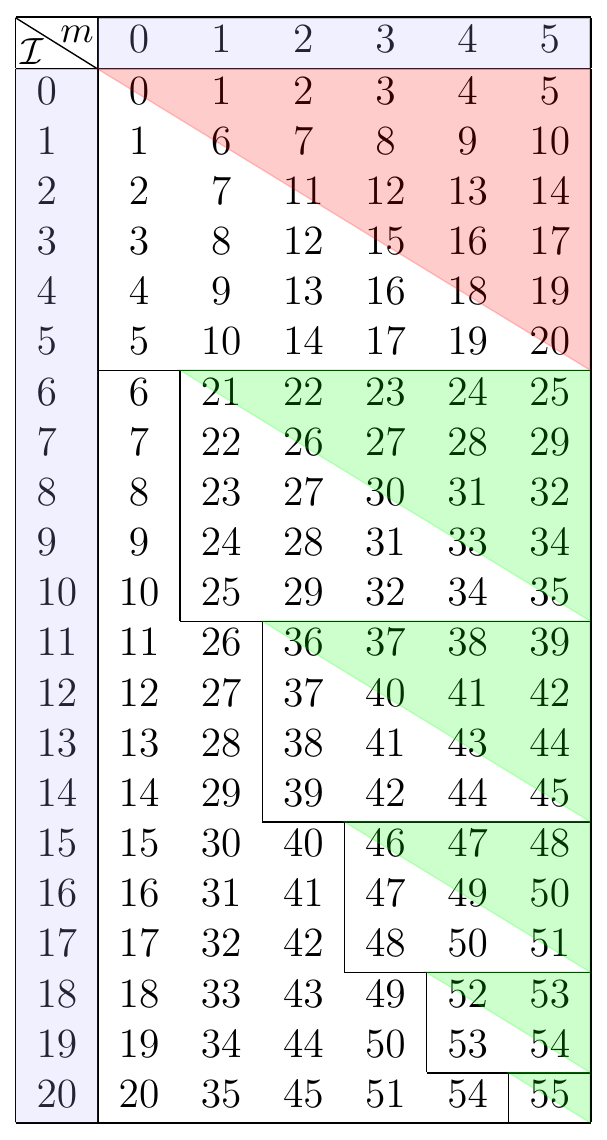}
\caption{  Map for N=3, M=5.
The map can be divided into two blocks, rows $0-5$ and $6-20$.
The latter can be generated with a recursive function of depth $5$.
Fig.~\ref{fig:map4} contains the additional blocks if we increase $N$ by $1$, i.e., $N=4$.
 \label{fig:map}
 }
\end{figure}

\begin{figure}[h]
\centering
\includegraphics[width=0.5\columnwidth]{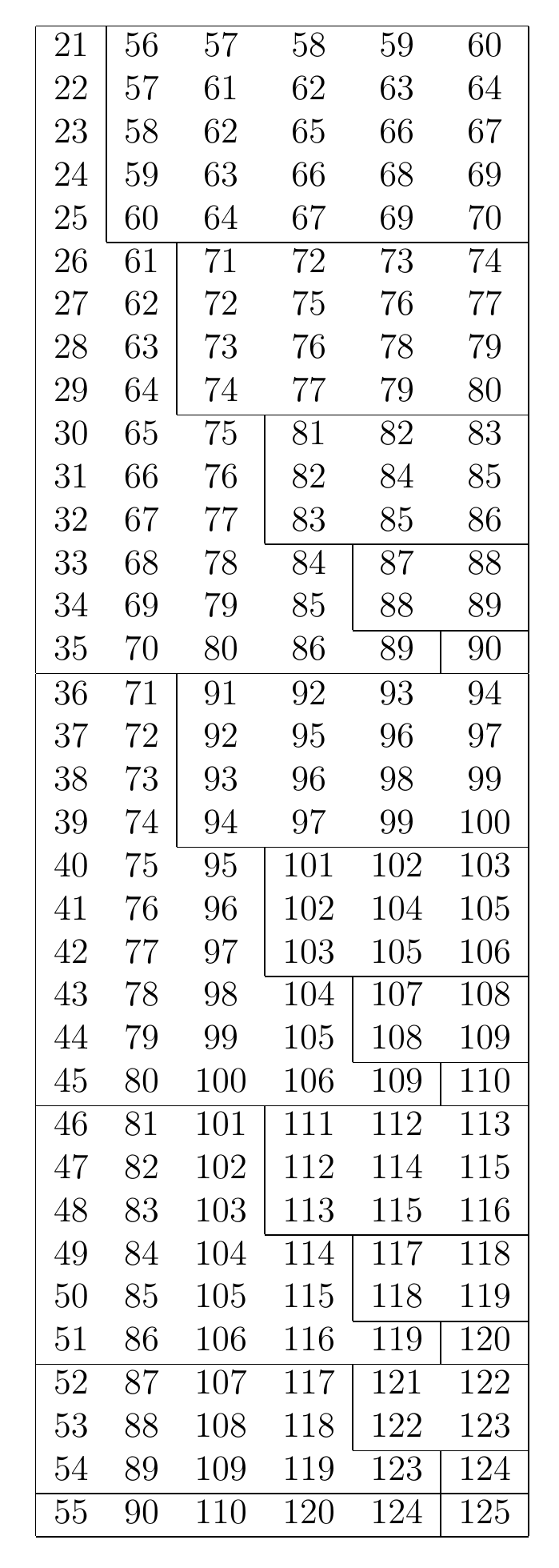}
\caption{ Additional blocks in the mapping for N=4 (and M=5),
with recursive depths $d=5,4,3,2,1$.
 \label{fig:map4}
 }
\end{figure}

\subsection{Recognising the pattern in the array $\maptoN$}

If we represent the mapping $\maptoN$ for a fixed M, say $M=5$, and a few values of $N$ (say $N=2,3,4,5$),
as $2d$ arrays with
the indices of $N-1$ mode states 
and occupation number of $N$th mode as the
 row and column indices, and the indices of $N$ mode states as array elements,
we observe a clear recursive pattern with some symmetries and well defined features. 
The map $\maptoN$ can be divided in blocks corresponding to addition of each site,
and each block consists of a series of square transpose-symmetric sub-blocks
 with sizes following a recursive pattern.

To illustrate this, the mapping for $N=3$ and $M=5$ is shown in Fig.\ref{fig:map}. 
Here, rows $0-5$ make complete map for $N=2$.
 and there is a series of smaller sub-blocks in the second block, i.e., rows $6-20$.
To generate this map, we can divide it into these two blocks. 
The first block is simple to generate.
 If we see the red shaded triangular region, 
 we find that the index simply starts from 0 on the top left, 
 increases one by one as we move to the right, and leaves $i$ column for the $ith$ row
  due to the condition that the occupation for the second site should not be less than that of the first one.
The pattern gets a little complicated in the second block, rows $6-20$. First, see the green shaded regions.
These are a set of triangles with smaller and smaller sizes, 
again due to the restriction on the occupation of the third site.
These shaded regions are the part of the map that can be recorded when the basis are formed.
Our main task is to calculate the unshaded part of the map.
The pattern that the unshaded regions follow is also easy to see, however. 
In the first block, it's simply the symmetric image of the upper triangular part.
In the second block, 
the lower triangular parts of the square blocks around the recursion of triangles follow the same rule. 
The one extra complication is the columns on the left side not included in the square regions, but, 
they also follow a regular pattern. 
The first column starting from the row $6$ continues the count of the first column from the first block
 until the end of the second block.
 The other columns starting from row $11,15,18,19,20$, follow the same rule.

The depth of the recursion for the triangles in the figure for $N=3$ is $d=5$.
For $N=4$, there are recursions of depth $5,4,3,2,1$ originating from this recursion of order $5$, see Fig.~\ref{fig:map4}.
This is a general feature of the mapping --- for any $N$, there are recursions of order $d,d-1,d-2,...,1$
 for each recursion of order $d$ in $N-1$ case!

\subsection{Generating the pattern}
\label{sec:genmap}
Generating the $N=2$ map is trivial. 
We will first describe here how the $N>2$ blocks can be generated
and then how we implement this in our code.
Starting with $N=3$ case shown in Fig.\ref{fig:map}, 
a recursive function can generate the second part of the map, i.e., rows $6-21$,
 taking $M=5$, the depth of the recursion $d=5$, initial index $i=21$, and the starting index of the column on the left $i_c = 6$ as inputs. 
For $N>3$, we have to generate recursions of lower order/depths that would require starting values for multiple columns on the left side, $M-d$ values for order $d$ recursion, to be specific. 
In summary,
we can divide the map into blocks corresponding to recursions of various orders
and a function can generate each block given the required starting indices 
and the recursion depth. 

 These blocks can be generated sequentially or in parallel, and combined with the block for two sites.
 The sequential implementation 
 is simpler, as all the arguments of the recursive function 
 ---
 starting index for the first triangle, the depth of the recursion, and 
 the set of starting indices for the leftover columns 
  --- are available at each call. 
  There are two points to consider for parallelisation. 
  First, the arguments of the recursive functions need to be calculated beforehand, and second, the workload needs to be distributed evenly between all processes.
The arguments are calculated sequentially, 
starting from the set for the first function call and, 
using the information on how much the indices are going to advance in that call, calculating the arguments for the next function call. 
To distribute the load evenly, 
the block sizes (heights, ($d(d+1)/2$ for a recursion of depth $d$) that every recursion would produce are to be calculated. 
Using this list, the total size among all processes can be evenly divided.
This way, different processes can have different number of function calls but the total workload determined by the total size of the map they produce is approximately the same.
Since, we only fill in a grid with indices that are obtained by either adding an integer to another or just plain ranges of integers,
 this method does not just scale linearly with the basis size, the prefactor or the slope of the scaling is also very small,
 which makes it computationally extraordinarily efficient.
 This means that the serial implementation can serve the purpose.

\section{FockMap library}
\label{sec:code}
The method presented in this article
is implemented in FockMap library. 
After a quick overview of the contents, 
the procedures contained in the library are briefly described.
An example test program that uses the library
 is explained at the end.

\subsection{Overview}
The procedures (subroutines and functions) that are meant to be called by 
the user program are
{\tt mapping, sizes, binomial, basis, basisall, ratios}, and {\tt ratiosall}.
All procedures are contained in separate files. 
The file name matches to the name of the procedure it contains.
Besides calculating the mapping, 
routines in this library can calculate
some other states related properties that could be required to compute the matrix elements of various operators.
See the description below for details.

\subsubsection{{\tt mapping}}
This subroutine implements the algorithm discussed in
sec.~\ref{sec:genmap} to
calculate the mapping $\maptoN$.
As described in sec.~\ref{sec:map},
the same array can be used 
for $\maptop$, $p=1,2,...,N$.

\subsubsection{{\tt basis} and {\tt basisall} }
 These routines calculate
 the number of permutations ${\cntset_{p}(k_{p})}$, 
 and the occupation frequencies ${f_{\nu_i}(\{\nu\})}$.
 {\tt basis} does it only for ${p=N-1,N}$ modes,
 whereas {\tt basisall} does it for ${p=0,1,...,N}$.
 
\subsubsection{{\tt ratios} and {\tt ratiosall} }
It also calculates
the ratios 
\begin{align*}
r_{\nu_i}(\{\nu^\prime\}) &\equiv \sqrt{ \cntset_{p-1}(j_{p-1})/\cntset_{p}(k_{p})}, \\
&= \sqrt{  (f_{\nu_i}(\{\nu^\prime\}) + 1)/p  },\\
j_{p-1} &\equiv \indset_{p-1}(\{\nu^\prime\}),\\
k_p &\equiv \indset_{p}(\{\nu\})
= \maptoN(\nu_i,j_{p-1}),
\end{align*}
which can be used 
instead of the bare numbers $\cntset_{p}(k_{p}), \cntset_{p-1}(j_{p-1})$
when calculating the overlap 
between permutation symmetric states with 
$p-1$ and $p$ modes.

\subsubsection{{\tt sizes}}
The subroutine {\tt sizes}
calculates the sizes of the permutation symmetric subspaces
for $p=0,1,...,N$ modes.
It also gives an array containing the shifts in the indexes 
if one is working with all these subspaces.

\subsubsection{{\tt binomial} }
This function simply calculates the binomial
$^nC_r$. It avoids large factorials as much as possible.

\subsection{Compiling FockMap}
\label{sec:comp}
FockMap uses a ``Makefile'' for the compilation. 
Edit the Makefile if you like to use a fortran compiler other than
{\tt gfortran} (we have tested gfortran only). FockMap can be compiled simply by running {\tt make} from the source directory on the {\it command line} on Unix or {\it terminal} on Mac.
This creates {\tt fockmap.a} archive that can be linked to the user programs wishing to exploit the permutation symmetry of their subsystems.

\subsection{Using FockMap}
\label{sec:use}

The use can best be illustrated by a test program
that calls various procedures in the FockMap library.
The {\tt test} program provided with the library 
(in the directory {\tt /fockmap/test/})
too can be compiled using the make command.
When executed,
it asks for the number of modes $N$ and the cutoff on the occupations $M$.
The source file for {\tt test} is {\tt /fockmap/test/main.f}, it illustrates the usage of the procedures in the FockMap. Here is a brief description of what it does.
It calls {\tt binomial} to calculate the permutation symmetric subspace size to allocate the memory to $\maptoN$ array
and calls {\tt mapping} to calculate the latter.
To obtain the occupation frequencies, number of permutations,
and the ratios described above,
 {\tt test} first calls {\tt sizes} to determine the sizes of the required arrays
and then calls {\tt basisall} and {\tt ratiosall}.
It prints $\maptoN$ and some other information on the standard output for inspection.

\subsection{Models' examples}
For a model to benefit from the FockMap library,
the only conditions that it needs to meet 
are, (i) it should contain one or more types of \emph{identical} subsystems whose permutations leave the model invariant, and (ii) we desire to solve the model for states and observables in the permutation symmetric subspace. 
To illustrate this point, here we list a few models describing the strong matter-light coupling.

\subsubsection{Dicke Model}
This model describes $N$ \emph{identical} two-level systems (2LSs) coupled collectively to a common cavity mode~\cite{Dicke}. 
The permutations over the 2LSs can be used
in this case.
$M=1$ for the 2LSs.

\subsubsection{Holstein-Tavis-Cummings model}
\label{sec:HTC}
As described in Ref.~\cite{zeb2020},
the permutations over the \emph{identical} vibrational modes of the organic molecule can be used.
$M$ will be equal to the cutoff on the vibrational states.
Including multiple intramolecular vibrational modes would require 
considering the permutational symmetry for each type.

\subsubsection{Spin-orbit coupling in organic microcavities}
Including the coupling between the singlet molecular excitons
and the triplet molecular excitons in organic microcavities
requires to consider three electronic levels per molecule (see Ref.~\cite{MartinezJCP2019}, for example).
The vibrational modes, if included in the model, 
can be treated just as in sec.~\ref{sec:HTC}.

\section{Conclusions}
We find that the mapping $\maptoN$ can be used to calculate
the matrix elements of various operators
when working in the permutation symmetric subspace.
We further find that $\maptoN$ has a fixed pattern that can be easily generated thus completely avoiding its calculation that scales badly.
This method is very efficient and has a negligible cost even for relatively large system sizes
that could allow studying a model in its thermodynamic limit.
A library
is provided that can give $\maptoN$ and some other quantities related to the permutation symmetric subspaces
for a given number of identical subsystems.

It is worth noting that, in practice, 
it is not important to know the reasons behind the characteristic pattern that the mapping $\maptoN$ gets. 
Even if we could find a mathematical expression for the elements of this mapping,
it would inevitably contain multiple sums with variable limits
and using it instead of the FockMap
would not be as efficient. 

Of course, calculations that must involve the 
rest of the Fock space cannot be done using the ideas presented in this work. However, in some cases, we can still use the permutation symmetry of fewer subsystems -- the hopping of charge careers in organic microcavities is an example~\cite{zeb2020}.
At the moment, the library works for a single set of subsystems over which the permutations are to be exploited.
Although, it is straightforward to use it and construct the \emph{global} indexing and maps when multiple sets or types are involved, 
this could be included in a future development. 
Further, to use this method and the library, 
the user has to frame their problem properly
and work out the relationship between the the mapping and the matrix elements of the operators they are interested in.
The examples in sec.~\ref{sec:matelem} are useful in this regard but
this task can still be made simpler if 
a future development of the method and the library
contains these relationships for commonly used types of operators.

\bibliographystyle{elsarticle-num}

\end{document}